\documentclass{iopart}

\usepackage{graphicx}
\usepackage{amssymb}
\usepackage{psfrag}

\begin{document}

\newcommand \be {\begin{equation}}
\newcommand \ee {\end{equation}}
\newcommand \bea {\begin{eqnarray}}
\newcommand \eea {\end{eqnarray}}

\title[]{Aging in the trap model as a relaxation further away from equilibrium}

\author{Eric Bertin}

\address{Universit\'e de Lyon, Laboratoire de Physique, ENS Lyon, CNRS,
46 All\'ee d'Italie, F-69007 Lyon}

\begin{abstract}
The aging regime of the trap model, observed for a temperature $T$ below
the glass transition temperature $T_\mathrm{g}$, is a prototypical example of non-stationary out-of-equilibrium state. We characterize this state by evaluating
its ``distance to equilibrium'', defined as
the Shannon entropy difference $\Delta S$ (in absolute value) between
the non-equilibrium state and the equilibrium state with the same energy.
We consider the time evolution of $\Delta S$ and show that, rather unexpectedly, $\Delta S(t)$ continuously increases in the aging regime, if the number of traps is infinite, meaning that the ``distance to equilibrium'' increases instead of decreasing in the relaxation process.
For a finite number $N$ of traps, $\Delta S(t)$ exhibits a maximum value before eventually converging to zero when equilibrium is reached. The time $t^*$ at which the maximum is reached however scales in a non-standard way
as $t^* \sim N^{T_\mathrm{g}/2T}$, while the equilibration time scales
as $\tau_\mathrm{eq} \sim N^{T_\mathrm{g}/T}$.
In addition, the curves $\Delta S(t)$ for different $N$ are found to rescale
as $\ln t/\ln t^*$, instead of the more familiar scaling $t/t^*$.
\end{abstract}

\pacs{02.50.-r,05.40.-a,64.70.P-}

\section{Introduction}

The aging phenomenon has attracted a lot of attention in the last decades,
both at the experimental and theoretical level \cite{Bouchaud-rev}.
Though being a genuine non-equilibrium state, the aging regime is
often intuitively considered as a progressive relaxation to the equilibrium
state, where the state of the system slowly gets
closer and closer to equilibrium \cite{Franz}.
In spite of the fact that most realistic models can only be studied through
extensive numerical simulations \cite{Kob},
some very simplified mean-field models,
like the trap model \cite{Bouchaud92,Monthus}
and the Barrat-M\'ezard model \cite{BM,Bertin-BM},
have been proposed in order to gain understanding on the time-dependent
probability distribution of microscopic configurations.
In such models, this time-dependent probability distribution can be
worked out exactly, thus providing a benchmark for testing possible
generic ideas or scenarios on the aging regime.
The first result that comes out of these simple models is that,
in the aging regime, microscopic configurations with small enough
trapping times are essentially equilibrated, while configurations with
very large trapping times are still strongly out of equilibrium.
The crossover trapping time between equilibrated and non-equilibrated
configurations is precisely of the order of the age of the system,
that is the time elapsed since it entered the low temperature phase.
As the system ages, the fraction of equilibrated configurations
increases, thus apparently confirming the scenario
that the system progressively gets closer and closer to equilibrium.

Although this scenario appears intuitively appealing, it would be interesting
to provide a quantitative characterization of this convergence to
equilibrium, for instance by computing a ``distance to equilibrium''
as a function of time.
A natural candidate to quantify the distance to equilibrium is the
Shannon entropy difference \cite{HatanoSasa,Martens}
between the non-equilibrium state and the equilibrium state with the
same average energy:
\be
\Delta S = S_\mathrm{eq}(\langle E\rangle_\mathrm{neq}) - S_\mathrm{neq}\geq 0
\ee
where $\langle E\rangle_\mathrm{neq}$ is the average energy in the non-equilibrium
state, and $S_\mathrm{neq}$ is the Shannon entropy of this state.
Indeed, the equilibrium state maximizes the entropy
for a given average energy, and this distance then vanishes by definition.
Note that this Shannon entropy difference identifies with the Kullback-Leibler
divergence \cite{Khinchin}, or relative entropy, between the corresponding
non-equilibrium and equilibrium states.
It is also interesting to note that $\Delta S$ characterizes
the dependence of the fluctuation-dissipation ratio on the observable
considered \cite{Martens}.

If an equilibrium state exists, $\Delta S$ should converge to
zero in the long time limit. It is however not obvious that the relaxation
to zero should be monotonous. To make a more specific statement,
we first note that for any stochastic markovian model having the canonical
distribution at temperature $T$ as equilibrium distribution,
the time-dependent free energy
$F(t)=\langle E\rangle_\mathrm{neq}(t)-TS_\mathrm{neq}(t)$
is a decreasing function of time \cite{vanKampen}.
On the other hand, the time-derivative of $\Delta S$ can be evaluated as
\be \label{dDS-dt}
\frac{d}{dt} \Delta S = \frac{1}{T(\langle E \rangle_\mathrm{neq})}
\, \frac{d}{dt} \langle E \rangle_\mathrm{neq}
- \frac{d}{dt} S_\mathrm{neq}
\ee
where we have introduced the microcanonical temperature $T(E)$, defined as
\be
\frac{1}{T(E)} = \frac{\partial S_{\mathrm{eq}}}{\partial E}.
\ee
Eq.~(\ref{dDS-dt}) can then be compared with the time derivative
of the free energy,
\be
\frac{dF}{dt} = T \left(  \frac{1}{T} \frac{d}{dt} \langle E\rangle_\mathrm{neq}
- \frac{d}{dt} S_\mathrm{neq} \right).
\ee
Hence, if $T(\langle E\rangle_\mathrm{neq})$ is close to the heat bath temperature
$T$,
\be
\frac{d}{dt}\Delta S \approx \frac{dF}{dt} \le 0,
\ee
and $\Delta S$ decreases. In the opposite situation,
if $T(\langle E\rangle_\mathrm{neq})$ is significantly different from the bath
temperature $T$, the evolution of $\Delta S$ with time may not be monotonous.

In addition, if an equilibrium state does not exist, as is the case
for instance in the trap model with an infinite number of traps
(in which case the equilibrium distribution is no longer normalizable),
no precise statement can a priori be made on the evolution of $\Delta S$.
However, the intuitive argument on the progressive equilibration of
the different degrees of freedom suggests that $\Delta S$ should
decrease with time.

In this short note, we compute the entropy difference as a function
of time in the trap model, and show that the behavior of this quantity
is quite different from naive expectations. Instead of monotonously
decreasing to zero with time, this quantity first increases during
the aging regime, before saturating and decaying to zero.
This non-monotonous behavior can be understood as the succession
of two regimes: a first aging regime which can be described
within a continuous formalism (meaning that the system essentially behaves
as if the number of traps was infinite), and a second regime where the finiteness
of the number of traps plays an important role.

\section{Trap model}

The trap model is defined as follows \cite{Bouchaud92,Monthus}.
A particle is trapped in one among a large number $N$ of traps,
whose bottom energy is drawn from an energy distribution $\rho(E)$, with $E<0$.
The standard choice for $\rho(E)$ is the exponential distribution
\be
\rho(E)= \frac{1}{T_\mathrm{g}}\, e^{-|E|/T_\mathrm{g}} \qquad (E<0),
\ee
which defines the energy scale $T_\mathrm{g}$.
The particle follows a continuous time markovian stochastic dynamics, with an escape rate $\Gamma_0 e^{E/T}$
from a trap of depth $|E|$, where $T$ is the heat bath temperature;
$\Gamma_0$ is a microscopic frequency, set to unity in the following.
After the particle escapes a trap, it chooses at random a new trap among
the $N$ traps. The transition rate $w_{ji}$ from trap $i$ to trap $j$ reads
\be
w_{ji} = \frac{1}{N}\, e^{E_i/T}.
\ee
This transition rate satisfies detailed balance with respect to the equilibrium Gibbs measure, namely
\be
w_{ji}\, e^{-E_i/T} = w_{ij}\, e^{-E_j/T}
\ee
which ensures that the probability distribution eventually reaches equilibrium.

In the limit of an infinite number of traps, all configurations with energy $E$
can be gathered in a single, coarse-grained, configuration
(see \cite{Monthus} for details), yielding for the transition rates
\be
W(E'|E) =  \rho(E')\, e^{E/T}.
\ee
The master equation for the probability $P(E,t)$ that a particle occupies a trap of energy $E$ at time $t$ is
\bea
\frac{\partial P}{\partial t}(E,t) = \int_{-\infty}^0
[W(E|E') P(E',t) - W(E'|E) P(E,t)]\, dE'.
\eea
Detailed balance is also satisfied within this continuous description, and the equilibrium measure reads
\be
P_\mathrm{eq}(E) = \frac{1}{Z}\, \rho(E)\, e^{-E/T}.
\ee
Quite importantly, it turns out that the normalization constant $Z$, defined as
\bea \nonumber
Z &=& \int_{-\infty}^0 \rho(E)\, e^{-E/T} \, dE\\
&=& \frac{1}{T_\mathrm{g}} \int_{-\infty}^0 \exp\left[\left(\frac{1}{T_\mathrm{g}}-\frac{1}{T}
\right)E\right] dE,
\eea
diverges for $T \le T_\mathrm{g}$.
This means that $P_\mathrm{eq}(E)$ becomes non-normalizable for $T \le T_\mathrm{g}$,
so that no equilibrium distribution can be defined in the limit
of an infinite number $N$ of traps where the continuous energy formalism
is valid. For a finite $N$, the equilibrium distribution however exists, and
concentrates on the few lowest energy levels.
For $T < T_\mathrm{g}$, the system enters an aging regime in which the probability distribution takes the form,
for large enough time $t$,
\be \label{travelling}
P(E,t) \approx \frac{1}{T}\, \Phi\left(\frac{E}{T}+\ln t\right),
\ee
where the dimensionless function $\Phi$ can be computed exactly \cite{Monthus}.
In the marginal case $T = T_\mathrm{g}$, the probability $P(E,t)$ has a slightly different form \cite{Bertin02}.
From Eq.~(\ref{travelling}), it is obvious that at large time the average energy
is given by $\langle E \rangle_\mathrm{ag} \approx -T\ln t$.
Computing the tails of $\Phi$, one finds that small traps with energies
much smaller than $\langle E \rangle_\mathrm{ag}$
in absolute value are essentially equilibrated, that is,
$P(E,t) \sim \rho(E)\,\exp(-E/T)$,
while large traps with $|E| \gg T\ln t$ remain equiprobable,
namely, $P(E,t) \sim \rho(E)$.
For a large but finite number of traps, this aging regime is interrupted when
$\langle E \rangle_\mathrm{ag}$ becomes of the order of the lowest energy levels.

\section{Entropy difference with the closest equilibrium state}

\subsection{Entropy in the aging regime}

Considering a finite number $N$ of traps, the time-dependent entropy $S(t)$ can be introduced
using the standard definition
\be
S(t) = - \sum_{i=1}^N P_i(t) \ln P_i(t).
\ee
where $P_i(t)$ is the probability that the particle is in trap $i$ at time $t$.
In the continuous energy approximation,
the number of traps having an energy between $E$ and $E+dE$ is $N\rho(E)dE$,
so that $P_i(t)$ can be estimated as
\be
P_i(t) \approx \frac{P(E_i,t)}{N\rho(E_i)}.
\ee
The entropy can then be written as
\be \label{entrop-continu}
S(t) = - \int_{-\infty}^0 P(E,t) \ln \left(\frac{P(E,t)}{N\rho(E)}\right) dE.
\ee
In the aging regime, one finds, using the ``travelling'' form 
Eq.~(\ref{travelling}) for $P(E,t)$
\be \label{Sag}
S_\mathrm{ag}(t) = \frac{1}{T_\mathrm{g}} \langle E \rangle_\mathrm{ag}
+ S_0 + \ln \frac{T}{T_\mathrm{g}} +\ln N,
\ee
where the constant $S_0$ is given by
\be
S_0 = -\int_{-\infty}^{\infty} \Phi(u) \ln \Phi(u)\, du.
\ee
For the temperature $T=T_\mathrm{g}/2$ used below in the numerical simulations,
one has $S_0 \approx 2.41$.

\subsection{$\Delta S$ in the infinite size limit}

We now wish to compute the entropy difference $\Delta S$ between the aging state
and the equilibrium state of the system with the same average energy.
To this aim, we need to determine both the equilibrium entropy and the
equilibrium average energy as a function of temperature, and then to express
the entropy as a function of the energy.

Let us first compute the equilibrium entropy at an arbitrary
temperature $T_\mathrm{a}$.
Taking $P_\mathrm{eq}(E,T_\mathrm{a}) = Z^{-1} \rho(E)\, e^{-E/T_\mathrm{a}}$, we find
for $T_\mathrm{a}>T_\mathrm{g}$,
using the definition of the entropy given in Eq.~(\ref{entrop-continu}), 
\be
S_\mathrm{eq}(T_\mathrm{a}) = \frac{1}{T_\mathrm{a}} \langle E \rangle_\mathrm{eq} +
\ln\left(\frac{T_\mathrm{a}}{T_\mathrm{a}-T_\mathrm{g}}\right)+\ln N.
\ee
In addition, the equilibrium average energy can be easily computed, yielding
\be
\langle E \rangle_\mathrm{eq}(T_\mathrm{a}) = -\frac{T_\mathrm{a}T_\mathrm{g}}{T_\mathrm{a}-T_\mathrm{g}} \qquad (T_\mathrm{a}>T_\mathrm{g}).
\ee
Inverting this relation to express the temperature as a function of the energy yields
\be
T_\mathrm{a}(E) = \left(\frac{1}{T_\mathrm{g}}+\frac{1}{E}\right)^{-1}.
\ee
Hence the equilibrium entropy reads, as a function of the energy
\be
S_\mathrm{eq}(E) = \frac{E}{T_\mathrm{g}} + \ln\left(\frac{|E|}{T_\mathrm{g}}\right) + 1+\ln N.
\ee
From Eq.~(\ref{Sag}), we can also express the entropy in the aging regime as a function of the energy $E \equiv \langle E \rangle_\mathrm{ag}$,
\be
S_\mathrm{ag}(E) = \frac{E}{T_\mathrm{g}} + S_0 + \ln \frac{T}{T_\mathrm{g}} +\ln N.
\ee
The entropy difference $\Delta S = S_\mathrm{eq}-S_\mathrm{ag}$ is thus obtained as
\be
\Delta S(E) = \ln\left(\frac{|E|}{T_\mathrm{g}}\right) -S_0
- \ln \frac{T}{T_\mathrm{g}}+1.
\ee
Using $\langle E\rangle_\mathrm{ag} \approx -T\ln t$ in the aging regime, one eventually finds
\be
\Delta S(t) \approx \ln\ln t - S_0 +1.
\ee
Hence $\Delta S$ increases as a function of time, which means that the
distribution $P(E,t)$ becomes more and more dissimilar to its equilibrium
counterpart. Thus, relaxation in the
aging regime of the trap model proceeds through states that get further
and further away from equilibrium.

\subsection{$\Delta S$ for a finite number of traps}

As long as the number of traps is infinite, no equilibrium state can be reached,
and $\Delta S(t)$ grows without bound. However, in the presence of a finite number of traps, the system
eventually reaches equilibrium, so that $\Delta S(t)$ should decrease to zero beyond a characteristic time scale.

\begin{figure}[t]
\includegraphics[width=0.5\columnwidth,clip]{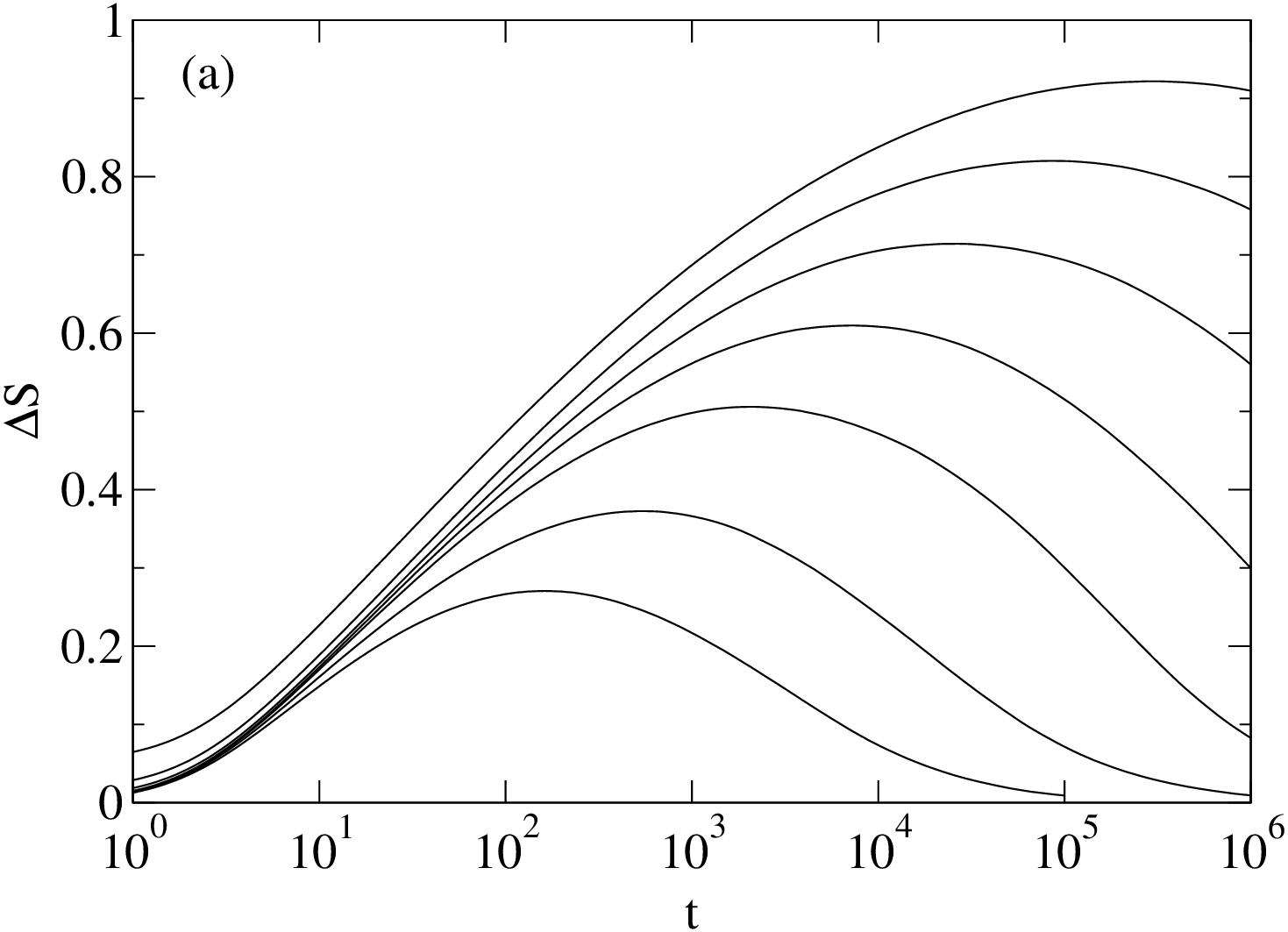}
\hfill
\includegraphics[width=0.5\columnwidth,clip]{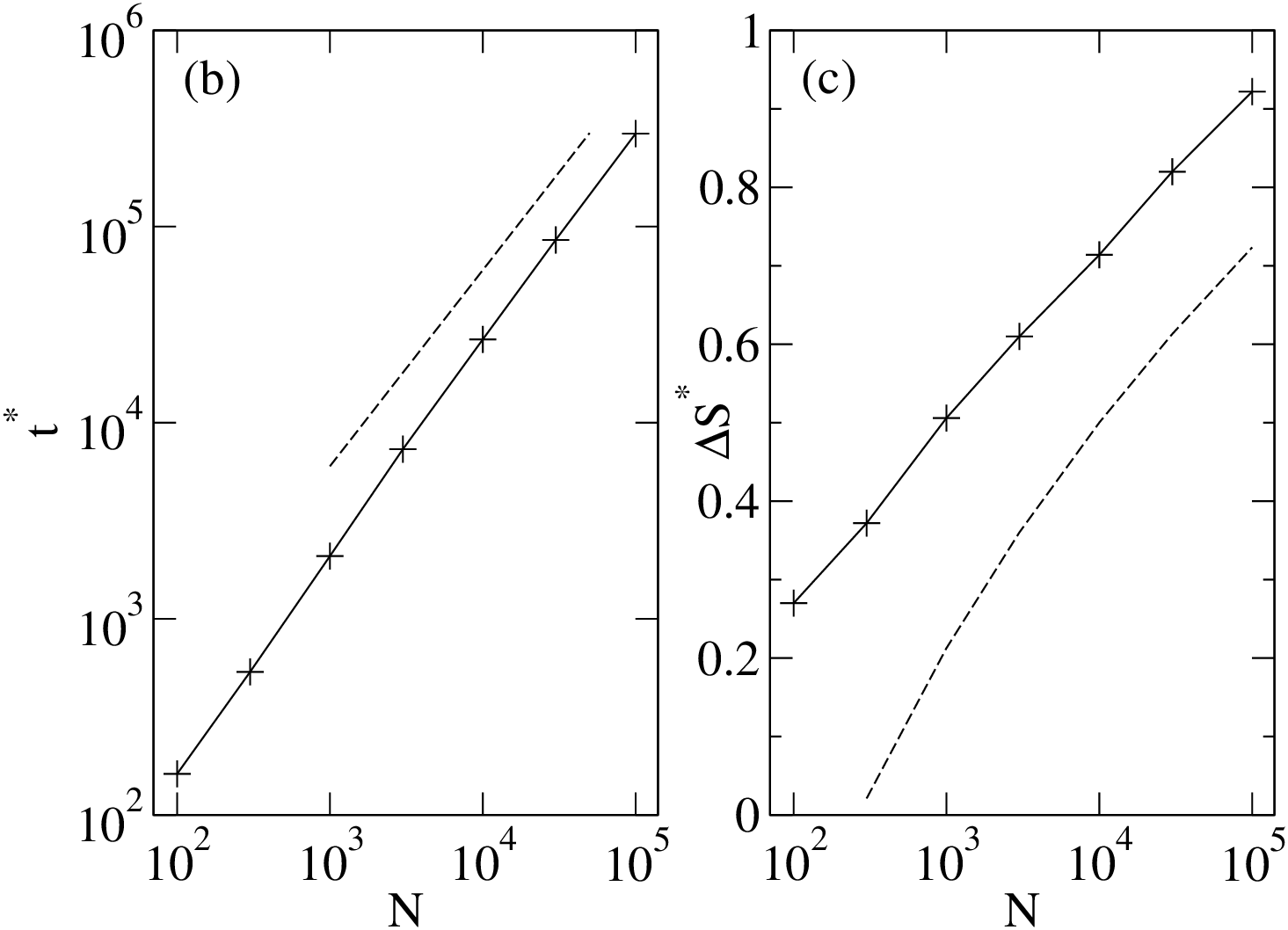}
\caption{(a) Entropy difference $\Delta S$ as a function of time,
for different
numbers of traps $N=10^2$, $3\times 10^2$, $10^3$, $3\times 10^3$, $10^4$,
$3\times 10^4$ and $10^5$ (from bottom to top).
Data were obtained by averaging over $500$ realizations of the disorder
(trap depths) the entropy difference computed for each sample ($T=0.5 T_\mathrm{g}$).
(b) Time $t^*$ at which $\Delta S$ is maximum, as a function of the number
$N$ of traps. The dashed line indicates a slope $T_\mathrm{g}/2T=1$.
(c) $\Delta S^* \equiv \Delta S(t^*)$ as a function of $N$.
The dashed line is the prediction for $\Delta S^*$ given
in Eq.~(\ref{eq-DSstar}).
}
\label{fig-DS}
\end{figure}

Fig.~\ref{fig-DS}(a) presents the numerical results obtained by simulating the stochastic dynamics of the model
with a finite number $N$ of trap. The resulting entropy difference $\Delta S(t)$ is averaged over
a large number of realizations of the disorder, that is, of the quenched energies of the traps.
One observes that $\Delta S(t)$ has a maximum value for a finite time
$t^*$, and then decreases to zero, albeit at a very slow rate.

A more surprising observation is the scaling with $N$ of $t^*$, for which we find, for $T=T_\mathrm{g}/2$, $t^* \sim N$ within numerical accuracy;
see Fig.~\ref{fig-DS}(b).
Indeed, a simple argument would to be say that the crossover
should be observed for a time $t^*$ of the order of the equilibration time
$\tau_\mathrm{eq}$, when the average energy $\langle E \rangle_\mathrm{ag}(t)$
becomes of the order of the minimal energy $E_\mathrm{min}$ of the $N$ traps.
Since trap energies are independent and identically distributed random variables
drawn from the exponential distribution $\rho(E)$, one finds
$E_\mathrm{min} \approx -T_\mathrm{g}\ln N$, yielding
$t^* \sim \tau_\mathrm{eq} \sim N^{T_\mathrm{g}/T}$.
For the temperature $T=T_\mathrm{g}/2$ considered in the simulations, one has
$\tau_\mathrm{eq} \sim N^2$, while the numerically observed scaling is
$t^* \sim N \ll \tau_\mathrm{eq}$. Hence the equilibration time alone
cannot account for the observed crossover.

As we shall see below, the reason for this discrepancy actually comes
from finite size effects in the calculation of the equilibrium entropy.
To take into account these finite size effects, one can as a first
approximation include a cut-off at $E_\mathrm{min} \approx -T_\mathrm{g}\ln N$
in the distribution $\rho(E)$.
Hence, for instance, the average equilibrium energy at temperature
$T_\mathrm{a}\equiv \beta_\mathrm{a}^{-1}$ is computed as
\be
\langle E \rangle_\mathrm{eq} = \frac{\beta_\mathrm{g}}{Z} \int_{-T_\mathrm{g} \ln N}^0 E \,
e^{(\beta_\mathrm{g}-\beta_\mathrm{a})E}\, dE
\ee
with $\beta_\mathrm{g} \equiv T_\mathrm{g}^{-1}$ and
\be \label{def-Z-finiteN}
Z = \beta_\mathrm{g} \int_{-T_\mathrm{g} \ln N}^0 e^{(\beta_\mathrm{g}-\beta_\mathrm{a})E}\, dE,
\ee
yielding
\be \label{eq-def-alpha}
\langle E \rangle_\mathrm{eq} = -\frac{T_\mathrm{g}}{\alpha} \left( 1 -\frac{\alpha \ln N}{N^{\alpha}-1} \right), \qquad \alpha \equiv 1-\beta_\mathrm{a} T_\mathrm{g}.
\ee
Interestingly, this expression of the energy takes a scaling form
with the number $N$ of traps, namely
\be \label{eq-Escal}
\langle E \rangle_\mathrm{eq} = T_\mathrm{g} \ln N f(\alpha \ln N),
\ee
where the function $f(x)$ is given by
\be
f(x) = \frac{1}{e^x -1} - \frac{1}{x} \qquad (x \ne 0).
\ee
It can be shown easily that $f(x) \sim -1/x$ for $x \to +\infty$ and
$f(x) \to -1/2$ when $x\to 0$, so that we set $f(0)=-1/2$.
One can also show that $f(x)$ is an increasing function, so that
the reciprocal function $g(y) \equiv f^{-1}(y)$ can be defined.
One then has
\be \label{alphaE}
\alpha(E) = \frac{1}{\ln N}\; g\left(\frac{E}{T_\mathrm{g} \ln N}\right).
\ee
From the behavior of $f(x)$, it follows that $g(y) \sim -1/y$ when $y \to 0$
and $g(-1/2)=0$. Note that the monotonicity of $g(y)$ implies that
it has a single zero.

Using Eq.~(\ref{entrop-continu}) and taking into account a finite-size cut-off
$-T_\mathrm{g} \ln N$, the equilibrium entropy for a finite number
$N$ of traps is found to be, once expressed as a function of the average
energy $E$,
\be
S_\mathrm{eq}(E) = \ln Z\big(\beta_\mathrm{a}(E)\big) + \beta_\mathrm{a}(E)\, E + \ln N,
\ee
where $\beta_\mathrm{a}(E)$ is obtained from Eq.~(\ref{alphaE}),
taking into account Eq.~(\ref{eq-def-alpha}).
Assuming, consistently with the numerical observations, that 
$t^* \ll \tau_\mathrm{eq}$, the non-equilibrium probability distribution
$P(E,t)$ should still take its aging form Eq.~(\ref{travelling}) 
for $t \sim t^*$. Hence the non-stationary entropy can still be
evaluated through Eq.~(\ref{Sag}) in this regime,
yielding for the entropy difference $\Delta S$
\bea \nonumber
\Delta S(E) &\equiv& S_\mathrm{eq}(E) - S_\mathrm{ag}(E)\\
&=& \ln Z\big(\beta_\mathrm{a}(E)\big)
+\big(\beta_\mathrm{a}(E) - \beta_\mathrm{g}\big) E
- \ln \frac{T}{T_\mathrm{g}} -S_0.
\label{eq-DS}
\eea
Considering $E=\langle E \rangle_\mathrm{ag}$ as a function of time,
we look for the time $t^*$ at which $\Delta S$ is maximum.
This maximum is obtained for
$d(\Delta S)/dt = [d(\Delta S)/dE]\,(dE/dt) = 0$,
and hence for an energy $E(t^*)$ such that $d(\Delta S)/dE=0$.
The derivative of $\Delta S$ with respect to $E$ reads
\be
\frac{d}{dE}\Delta S = \beta_\mathrm{a}(E) - \beta_\mathrm{g} = -\beta_\mathrm{g} \, \alpha(E),
\ee
where we have used the relation $d\ln Z/d\beta_\mathrm{a}=-E$.
The maximum of $\Delta S$ is thus obtained for $t=t^*$ such
that $\alpha(E(t^*))=0$.
From Eq.~(\ref{alphaE}), this corresponds to $E(t^*)=-\frac{1}{2} T_\mathrm{g} \ln N$.
Using $E(t) \sim -T\ln t$, we eventually obtain
\be \label{eq-tstar}
t^* \sim N^{T_\mathrm{g}/2T}.
\ee
This estimation is indeed consistent with the numerical observations,
as seen on Fig.~\ref{fig-DS}(b).

Let us now evaluate $\Delta S^* \equiv \Delta S(t^*)$.
From Eq.~(\ref{def-Z-finiteN}), one can rewrite $Z$ as
$Z=\ln N\, h(\alpha \ln N)$, where the function $h(x)$ is defined by
\be
h(x) = \frac{1}{x} (1-e^{-x}), \qquad x \ne 0,
\ee
and $h(0)=1$. Then, using Eq.~(\ref{eq-DS}) and (\ref{eq-Escal}),
$\Delta S$ can be rewritten as
\be \label{eq-DS2}
\Delta S = \ln \ln N - S_1 
+ \ln h(\alpha \ln N) - \alpha \ln N f(\alpha \ln N),
\ee
where we have introduced the notation $S_1\equiv S_0+\ln(T/T_\mathrm{g})$.
For $t=t^*$, one has $\alpha=\alpha(E^*)=0$, so that
\be \label{eq-DSstar}
\Delta S^* = \ln \ln N -S_1.
\ee
This result qualitatively agrees with the numerical simulations,
though a significant shift is observed, as seen on Fig.~\ref{fig-DS}(c).
This shift is likely to be due to
the approximation made by taking into account the finite $N$ effects
through a simple cut-off in the energy distribution.

\begin{figure}[t]
\begin{center}
\includegraphics[width=0.5\columnwidth,clip]{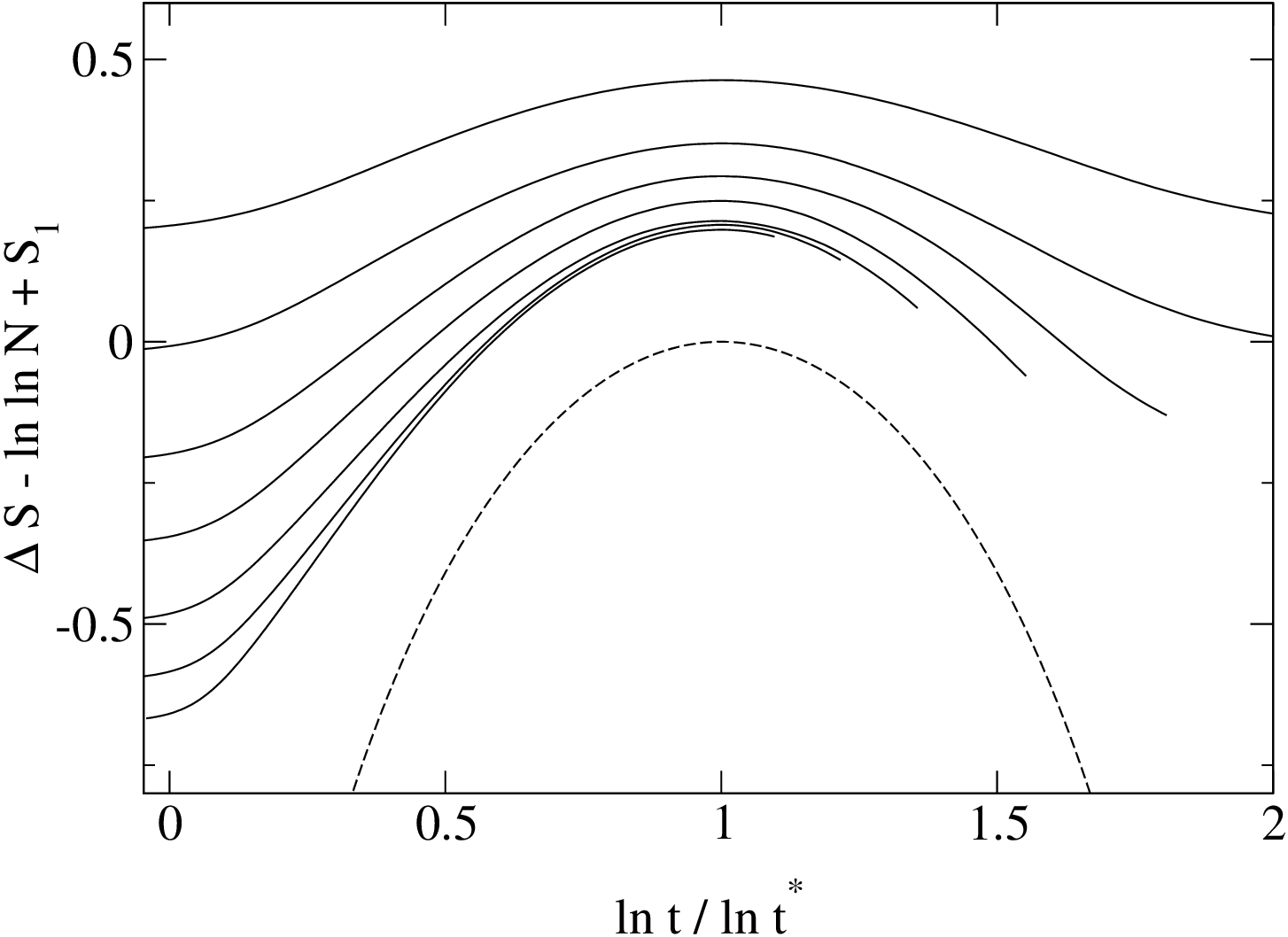}
\caption{Shifted entropy difference $\Delta S -\ln \ln N + S_1$
as a function of the rescaled logarithmic time $\ln t/\ln t^*$,
for different numbers of traps.
Data are the same as in Fig.~\ref{fig-DS}(a);
from top to bottom: $N=10^2$, $3\times 10^2$, $10^3$, $3\times 10^3$, $10^4$,
$3\times 10^4$ and $10^5$.
The dashed line corresponds to the analytical form given in Eq.~(\ref{eq-DS4}).
}
\label{fig-DS2}
\end{center}
\end{figure}

In addition, coming back to Eq.~(\ref{eq-DS2}), we can express
$\alpha \ln N$ as a function of the rescaled energy
$E/(T_\mathrm{g} \ln N)$ through Eq.~(\ref{alphaE}), and then use
the time-dependence $E=\langle E \rangle_{\mathrm{ag}} \approx -T\ln t$
for the energy, yielding
\be \label{eq-ETg}
\frac{E}{T_\mathrm{g} \ln N} = -\frac{1}{2}\, \frac{\ln t}{\ln t^*}
\ee
where we have also used the expression (\ref{eq-tstar}) of $t^*$.
Combining Eqs.~(\ref{eq-DS2}), (\ref{alphaE}) and (\ref{eq-ETg}),
one finally finds that
$\Delta S - \ln \ln N +S_1$ is a function of the logarithmically
rescaled time $\ln t/\ln t^*$:
\be \label{eq-DS4}
\Delta S - \ln \ln N +S_1 = \ln h(x) - xf(x), \quad
x \equiv g\left(-\frac{\ln t}{2\ln t^*}\right).
\ee
We test this property on Fig.~\ref{fig-DS2}, and find
that, though strong finite size effects are present,
this rescaling of the numerical data seems to be asymptotically satisfied.
The expression given in Eq.~(\ref{eq-DS4}) is also plotted
on Fig.~\ref{fig-DS2} for comparison.
Here again, a significant shift on $\Delta S$ is present with respect
to the numerical data, for the same reasons as on Fig.~\ref{fig-DS}(c).
However, apart from this shift, the overall shape is reasonably well
reproduced, except in the tails. These discrepancies for both
short and large times are due to the fact that the travelling form
Eq.~(\ref{travelling}) of the aging distribution $P(E,t)$
is no longer valid in these regimes.

\section{Discussion and conclusion}

In this short note, we have computed in the trap model
the entropy difference $\Delta S$ between
the aging state and the equilibrium state with the same energy.
For an infinite number of traps, a simple calculation shows that $\Delta S$
actually increases without bound as time elapses, contrary to the naive
expectation based on a scenario of progressive equilibration.
For a finite number of traps, when an equilibrium state exists for a heat
bath temperature $T<T_\mathrm{g}$, $\Delta S$ first increases before eventually decreasing
to zero in the long time limit. The characteristic time $t^*$ at which the maximum
of $\Delta S$ occurs is however much smaller than the equilibration time
$\tau_{\mathrm{eq}}$, and one actually finds a strong scale separation between
these two times, according to $t^* \sim (\tau_{\mathrm{eq}})^{1/2}$.

Beyond these results, let us mention that the entropy difference $\Delta S$
appears as an interesting quantity to statistically characterize
the aging regime. Indeed, the standard Kovacs memory effect \cite{Kovacs},
reproduced in numerous numerical or analytical models
\cite{Berthier,Mossa,Bertin-Kov,Buhot,Cugliandolo,Brey},
confirms that standard macroscopic observables like the energy or the density
are not enough to characterize the macroscopic state of a system in the
glassy phase. The Kovacs protocol consists in suddenly raising the temperature
from a value $T_1$ (at which the system is glassy) to a value $T_2>T_1$,
precisely at the time $t_\mathrm{r}$ when a given macroscopic observable
(typically density or energy) reaches its equilibrium value at $T_2$
\footnote{A correction related to the rapid energy increase of
fast degrees of freedom however has to be
taken into account in realistic models. In the trap model, these fast degrees
of freedom, like bottom-of-the-well vibrations, are not present.}.
If such an observable was enough to describe the macroscopic state
of the system, this observable would no longer evolve with time
for $t>t_\mathrm{r}$, having already reached its
equilibrium value at temperature $T_2$. In contrast, observations
show a non-monotonic evolution of the observable, which starts to depart
from its equilibrium value before eventually returning to it.
At a macroscopic level, this behavior can only be understood if at least
another variable, that has not yet reached its equilibrium value, is present.
As such a variable should in some sense quantify the deviation from equilibrium,
the entropy difference $\Delta S$ turns out to be a natural candidate.
In addition, as the energy is close to the equilibrium energy at $T_2$,
the relaxation of $\Delta S$ should be essentially monotonous,
according to the arguments presented in the introduction.

\subsection*{  }

\end{document}